\documentclass[pra,aps,twocolumn,10pt]{revtex4-2}
\usepackage[english]{babel}
\usepackage{amsmath,amssymb,amsfonts,bbold}

\usepackage{graphicx}
\usepackage[colorlinks=True,linkcolor=blue,citecolor=blue,urlcolor=blue]{hyperref}

\newcommand{\ii}{\mathrm{i}}

\DeclareMathOperator{\tr}{\mathrm{tr}}

\usepackage{enumitem}
\setlist[enumerate]{align=parleft,left=0pt..2em}
\setlist[itemize]{align=parleft,left=0pt..1em}

\makeatletter
\renewcommand*{\fnum@figure}{{\normalfont\bfseries \figurename~\thefigure}}
\makeatother

\begin{document}
\title{Exceptional, but Separate: Precursors to Spontaneous Symmetry Breaking}
\author{Lewis Hill$^{1,\dagger,a}$}
\author{Julius T. Gohsrich$^{1,2,\dagger,b}$}
\author{Alekhya Ghosh$^{1,2}$}
\author{Jacob Fauman$^{1,2}$}
\author{Pascal Del'Haye$^{1,2}$}
\author{Flore K. Kunst$^{1,2}$}
\affiliation{$^{1}$Max Planck Institute for the Science of Light, 91058 Erlangen, Germany \\$^{2}$Department of Physics, Friedrich-Alexander Universität Erlangen-Nürnberg, 91058 Erlangen, Germany \\$^\dagger$\emph{Contributed equally and should be considered as co-first authors.}\\$^a$\href{mailto:lewis.hill@mpl.mpg.de}{\emph{lewis.hill@mpl.mpg.de}} \qquad $^b$\href{mailto:julius.gohsrich@mpl.mpg.de}{\emph{julius.gohsrich@mpl.mpg.de}}}
\date{\today}
\begin{abstract}
Spontaneous symmetry breaking (SSB) and exceptional points (EPs) are often assumed to be inherently linked. Here we investigate the intricate relationship between SSB and specific classes of EPs across three distinct, real-world scenarios in nonlinear optics. In these systems, the two phenomena do not coincide for all classes of EPs; they can occur at dislocated points in parameter space. This recurring behavior across disparate platforms implies that such decoupling is not unique to these optical systems, but likely reflects a more general principle. Our results highlight the need for careful analysis of assumed correlations between SSB and EPs in both theoretical and applied contexts. They deepen our understanding of nonlinear dynamics in optical systems and prompt a broader reconsideration of contexts where EPs and SSB are thought to be interdependent.
\end{abstract}
\maketitle
\section{\label{sec:Introduction}Introduction}
\begin{figure}
    \centering
    \includegraphics[width=\linewidth]{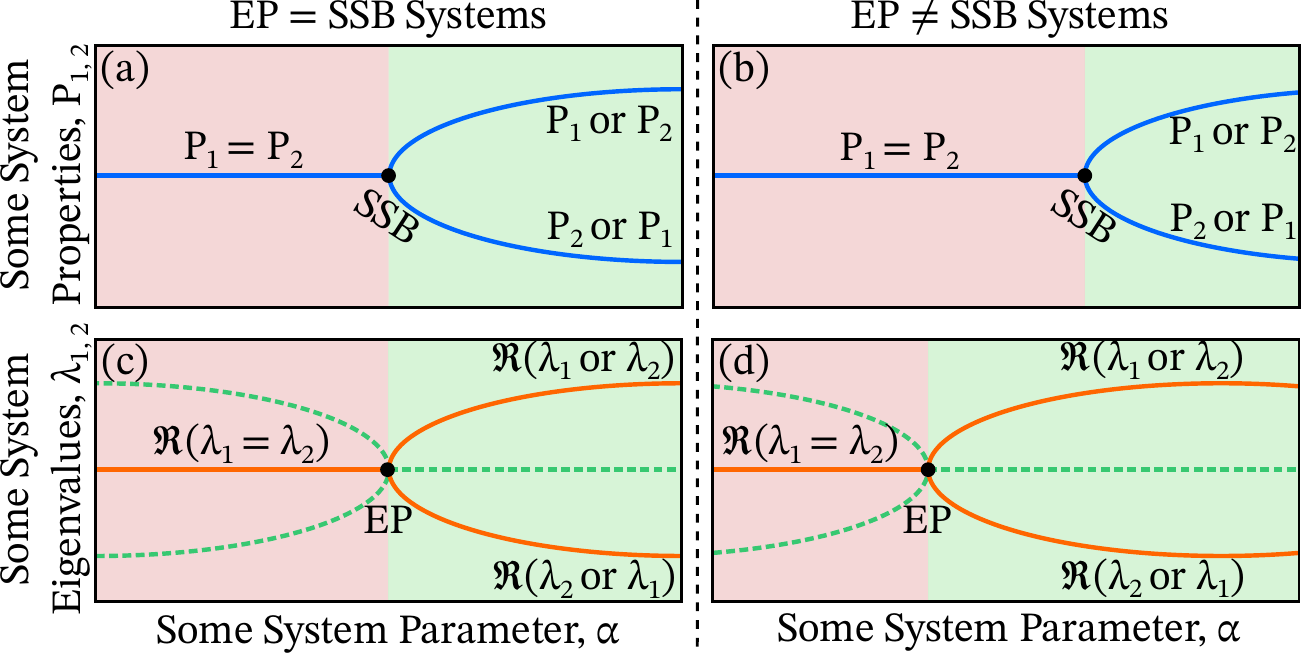}
    \caption{\textbf{Illustrative example systems in which an exceptional point (EP) and spontaneous symmetry breaking (SSB) either coincide (left) or are dislocated (right).} Panels (a,b) show the SSB of two system properties, $P_1$ and $P_2$, occurring where the relation $P_1=P_2$ abruptly breaks as a system parameter, $\alpha$, is varied. Panels (c,d) depict EPs along $\alpha$, where two eigenvalues, $\lambda_1$ and $\lambda_2$ become degenerate and their associated eigenvectors coalesce. While SSB is often assumed to coincide with an EP, as in panels (a,c), this is not guaranteed for all types of EPs: panels (b,d) illustrate that the SSB and an EP -- such as one derived from the system Jacobian -- can occur dislocated in parameter space, as we exemplify in this work.}
    \label{fig:EP_Introduction}
\end{figure}

The spontaneous symmetry breaking (SSB) of two system properties, $P_1$ and $P_2$ (Fig.~\ref{fig:EP_Introduction}(a,b)), is often regarded as co-occurring with an exceptional point (EP)~\cite{wang_petermann-factor_2020,wang_steering_2020,bergman_observation_2021} -- a degeneracy of eigenvalues (Fig.~\ref{fig:EP_Introduction}(c,d)) -- which arises exclusively in non-Hermitian systems~\cite{bergholtz_exceptional_2021,kato_perturbation_1995,berry_physics_2004,heiss_physics_2012}. Here, we use the term “co-occurring” to mean that the onset of SSB and the emergence of an EP coincide at the same location in parameter space (Fig.~\ref{fig:EP_Introduction}(a,c)), as opposed to being dislocated (Fig.~\ref{fig:EP_Introduction}(b,d)). In this work, we present the stark latter possibility: a decoupling of these phenomena, observed not in a single case but across three distinct, real-world scenarios in nonlinear optics. This separation arises from the interplay between system-level SSB and EPs derived from the eigenvalues of the system's Jacobian, underscoring the critical importance of identifying the appropriate origin of EPs when predicting the onset of SSB.

The SSB of light in Kerr resonators has become a vibrant and active area of research within nonlinear optics. Over the past decade and beyond, investigations into this phenomenon have yielded numerous significant findings~\cite{kaplan_enhancement_1981,kaplan_directionally_1982,wright_theory_1985,geddes_polarisation_1994,xu_experimental_2014,li_enhanced_2015,rossi_spontaneous_2016,del_bino_symmetry_2017,cao_experimental_2017,woodley_universal_2018,hendry_spontaneous_2018,bino_microresonator_2018,copie_interplay_2019,wu_symmetry-breaking-induced_2019,cuong_spontaneous_2019,hill_effects_2020,garbin_asymmetric_2020,dolinina_spontaneous_2020,moroney_logic_2020,ghalanos_kerr-nonlinearity-induced_2020,woodley_self-switching_2021,silver_critical_2021,xu_spontaneous_2021,garbin_dissipative_2021,fatome_self-symmetrization_2021,li_microcavity_2021,silver_nonlinear_2021,xu_breathing_2022,campbell_counterpropagating_2022,mai_nonreciprocal_2022,mai_double_2022,moroney_kerr_2022,hill_multi-stage_2023-1,ghosh_four-field_2023,bitha_bifurcation_2023,wang_spontaneous_2023,quinn_random_2023,campbell_dark_2023,mai_spontaneous_2024,lucas_polarization_2024,campbell_frequency_2024,ghosh_controlled_2024,hill_symmetry_2024,coen_nonlinear_2024,pal_linear_2024,rah_demonstration_2024,ghosh_phase_2024,zhang_integrated_2025}, and have enabled a wide range of applications, including enhanced rotation sensors~\cite{kaplan_enhancement_1981,kaplan_directionally_1982,wright_theory_1985}, gyroscopes~\cite{silver_nonlinear_2021}, isolators and circulators~\cite{bino_microresonator_2018}, all-optical logic gates~\cite{moroney_logic_2020}, dual-frequency resonator coupling~\cite{ghalanos_kerr-nonlinearity-induced_2020}, and the generation of vectorial temporal cavity solitons~\cite{xu_spontaneous_2021}. It has also driven advances in vectorial frequency comb enhancement via self-crystallization dynamics~\cite{campbell_dark_2023-1,campbell_frequency_2024}, polarization control~\cite{moroney_kerr_2022}, random number generation~\cite{quinn_random_2023}, novel temporal structures such as faticons~\cite{lucas_polarization_2024}, polarization conformity in symmetry-broken soliton chains~\cite{hill_symmetry_2024}, and controlled light routing in coupled-resonator optical waveguides (CROWs)~\cite{ghosh_controlled_2024}.

Since SSB arises from an instability in the system dynamics -- and since the stability is governed by the properties of the Jacobian -- it is natural to ask whether, and where, the Jacobian exhibits EPs. In addressing this question, we uncover the scenario illustrated in Fig.~\ref{fig:EP_Introduction}(b,d), where SSB and Jacobian EPs occur at distinctly different locations in parameter space -- i.e. they are dislocated.

This cautionary insight is relevant across all areas of physics and mathematics that investigate the relationship between SSB and EPs. Accordingly, we begin by introducing SSB in the context of Kerr resonators from first principles, in the form of a focused research review of the most relevant systems, with the specific aim of guiding readers from outside the field of nonlinear optics.
\begin{figure*}
    \centering
    \includegraphics[width=0.8\textwidth]{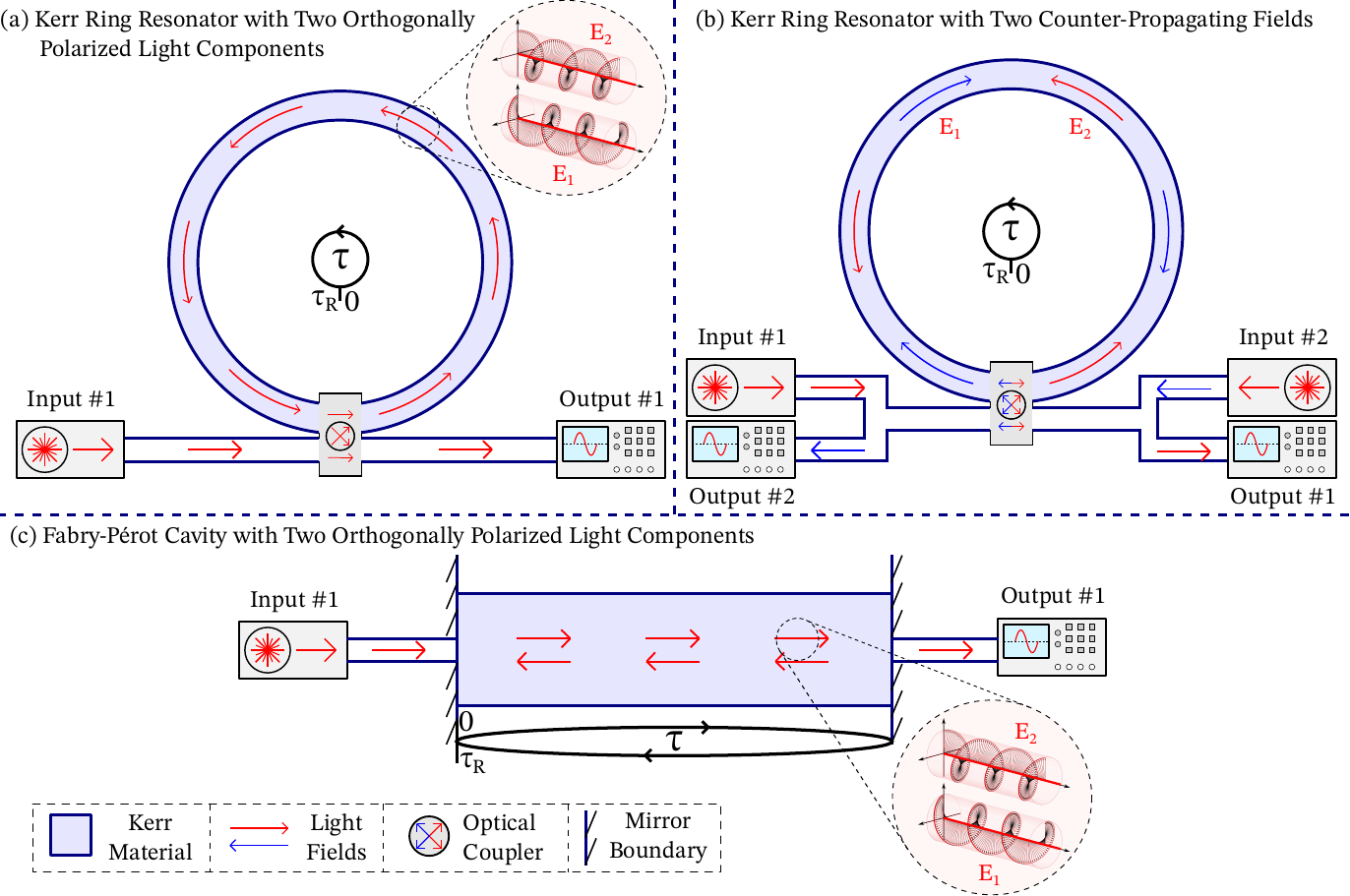}
    \caption{\textbf{Kerr resonator schematics.} (a) Two co-propagating light field components, $E_{1,2}$ -- with left- and right-circular polarizations, respectively -- are coupled into a Kerr ring resonator using a single linearly polarized input pump. (b) Two identical input pumps introduce counter-propagating light fields, $E_{1,2}$, into the Kerr ring resonator. (c) In a Fabry-Pérot cavity, reflections at the cavity boundaries cause rebounded fields to coexist, still true when multiple polarizations $E_{1,2}$ are present. In all cases, $\tau$ denotes the fast-time axis used for modelling, which ranges from $0$ to $\tau_\mathrm{R}$, the round-trip time.}
    \label{fig:Schematics}
\end{figure*}
\subsection{\label{sec:KerrRingIntro}An Introduction to Kerr Resonators}
Figure~\ref{fig:Schematics} illustrates two of the most fundamental optical resonator configurations: the ring (Fig.~\ref{fig:Schematics}(a,b)) and the Fabry-Pérot resonator (Fig.~\ref{fig:Schematics}(c)). In the ring configuration, and in the absence of backscattering, a single injected light field can circulate without encountering its own reflection. In contrast, in a Fabry-Pérot cavity, even a single injected field will inevitably reflect off the cavity boundaries. As a result, these two configurations are typically modeled using slightly different approaches.

When light interacts with matter, the initially unpolarized atoms within the medium can become electrically polarized. Assuming a dielectric material that is lossless, isotropic, and dispersionless, the polarization density $P$ can be expressed as:
\begin{equation}\label{Eq:PolDen}P=\varepsilon_0\chi^{(1)}E+\varepsilon_0\chi^{(2)}EE+\varepsilon_0\chi^{(3)}EEE+\ldots \,,
\end{equation}
where $\varepsilon_0$ is the vacuum permittivity, $E$ is the electric field of the incident light, and $\chi^{(n)}$ are the material's $n$th-order electric susceptibility constants. The term with~$\chi^{(1)}$ describes the linear response of the material, while the higher-order terms account for its nonlinear response.

In Kerr materials, where second- and higher-than-third-order susceptibilities are negligible, the polarization density simplifies to
\begin{equation}\label{Eq:KerrPolDen}
    P=\epsilon_0\chi^{(1)}E+\epsilon_0\chi^{(3)}EEE.
\end{equation}
\subsubsection{The Lugiato-Lefever Equation}
One of the most successful models for Kerr ring resonators is the Lugiato-Lefever equation (LLE) \cite{lugiato_spatial_1987,lugiato_lugiatolefever_2018}, which in its purely temporal \cite{haelterman_dissipative_1992} and normalized form is given by
\begin{equation}\label{Eq:LLE}
    \frac{\partial E}{\partial t} = E_{\mathrm{in}} - E - \ii\theta E - \ii\eta\frac{\partial^2 E}{\partial\tau^2} + \ii|E|^2E.
\end{equation}
The LLE describes the evolution of the complex envelope of the intracavity electric field, $E$, over a slow timescale~$t$, typically on the order of the photon lifetime in the cavity. The terms $E_{\mathrm{in}}$ and $-E$ account for, respectively, the continuous injection and loss of light from the cavity. The term $-\ii\theta E$ models the phase shift due to detuning, where $\theta$ is the normalized frequency mismatch between the input pump and the nearest cavity resonance. The term $-\ii\eta\partial^2_\tau E$ accounts for group-velocity dispersion (GVD) along the so-called fast-time axis $\tau$, which spans one round-trip of the cavity and is bounded between 0 and the round-trip time $\tau_\mathrm{R}$, as illustrated in Fig.~\ref{fig:Schematics}. The parameter $\eta = \pm 1$ denotes normal ($+1$) or anomalous ($-1$) GVD. The final term, $\ii |E|^2E$, represents self-phase modulation (SPM), a nonlinear phase shift arising from the Kerr interaction between the circulating field and the resonator material.

The LLE has been adapted to model Fabry-Pérot cavities~\cite{cole_theory_2018,campbell_dark_2023-1}, where counter-propagating field components must be taken into account due to the inherent reflections at the cavity boundaries, as previously discussed. This is achieved by introducing an additional term involving a round-trip average of the field intensity, denoted by~$\langle \cdot \rangle$~\cite{cole_theory_2018}. The resulting Fabry-Pérot LLE is
\begin{equation}\label{Eq:LLEFP}
    \frac{\partial E}{\partial t} = E_{\mathrm{in}} - E - \ii\theta E - \ii\eta\frac{\partial^2 E}{\partial\tau^2} + \ii|E|^2E + 2\ii\langle |E|^2\rangle E.
\end{equation}
The denormalization of both LLE models to experimental parameters is provided in the Methods section.
\subsection{\label{sec:CLLEIntro}Coupled Lugiato-Lefever Equations}
So far, we have considered resonators supporting a single circulating field envelope. However, Fig.~\ref{fig:Schematics} shows three configurations in which two field envelopes are coupled. In such cases, cross-phase modulation (XPM) must also be accounted for: the field envelope $E_1$ affects $E_2$, and vice versa. These three coupled-field setups are:
\begin{itemize}
    \item \textbf{A Kerr ring resonator with a single, linearly polarized input pump} (Fig.~\ref{fig:Schematics}(a)). We describe two co-propagating field components in terms of their left- and right-circular polarizations, $E_{1,2}$, ~\cite{geddes_polarisation_1994} governed by
        \begin{multline}\label{Eq:RCircLLE}
            \frac{\partial E_{1,2}}{\partial t} = E_{\mathrm{in}} - E_{1,2} - \ii\theta E_{1,2} - \ii\eta\frac{\partial^2 E_{1,2}}{\partial\tau^2} \\ + \ii A|E_{1,2}|^2E_{1,2}+\ii B|E_{2,1}|^2E_{1,2}.
        \end{multline}
    \item \textbf{A Kerr ring resonator with two identical counter-propagating pumps} (same input frequency and intensity), introducing two counter-propagating fields, $E_{1,2}$ (Fig.~\ref{fig:Schematics}(b)). The coupled LLE now is~\cite{campbell_counterpropagating_2022}
\begin{multline}\label{Eq:RCountLLE}
    \frac{\partial E_{1,2}}{\partial t} = E_{\mathrm{in}} - E_{1,2} - \ii\theta E_{1,2} - \ii\eta\frac{\partial^2 E_{1,2}}{\partial\tau^2}\\ + \ii A|E_{1,2}|^2E_{1,2}+\ii B\langle|E_{2,1}|^2\rangle E_{1,2}.
\end{multline}
    \item \textbf{A Kerr Fabry-Pérot cavity with linearly polarized input light} (Fig.~\ref{fig:Schematics}(c)). The field is decomposed into left- and right-circularly polarized co-propagating components, $E_{1,2}$~\cite{hill_symmetry_2024,campbell_dark_2023}, governed by
\begin{align}\label{Eq:FPCircLLE}
    \frac{\partial E_{1,2}}{\partial t} = E_{\mathrm{in}} &- E_{1,2} - \ii\theta E_{1,2} - \ii \eta \frac{\partial^2E_{1,2}}{\partial\tau^2} + \ii A|E_{1,2}|^2E_{1,2} \notag \\
    &+ \ii B|E_{2,1}|^2E_{1,2} + 2\ii A\langle|E_{1,2}|^2 \rangle E_{1,2} \notag \\
    &+ \ii B\langle |E_{2,1}|^2\rangle E_{1,2}  + \ii B\langle E_{1,2} E_{2,1}^*\rangle E_{2,1}.
\end{align}
\end{itemize}
In all cases, $A$ and $B$ are constants that define the relative strengths of SPM and XPM, respectively. A detailed discussion of the possible values these constants can practically take is provided in Ref.~\citenum{hill_effects_2020}.

Equations~(\ref{Eq:RCircLLE}-\ref{Eq:FPCircLLE}) differ significantly in their fast-time~\mbox{($\tau$)} dynamics. They exhibit varying susceptibilities to -- and symmetry breaking of -- turning patterns, bright and dark temporal cavity solitons~\cite{xu_spontaneous_2021,campbell_dark_2023-1,hill_symmetry_2024}, breathers~\cite{xu_breathing_2022}, faticons~\cite{lucas_polarization_2024}, soliton self-crystallization dynamics~\cite{campbell_frequency_2024,campbell_dark_2023-1}, and polarization conformity in soliton chains~\cite{hill_symmetry_2024}.

However, in this work we focus on \emph{homogeneous} states, i.e., states that are constant in $\tau$. This assumption simplifies the equations in two key ways: (i) the fast-time derivative vanishes, $\partial^2_\tau E_{1,2} = 0$, and (ii) fast-time averaging becomes trivial, such that $\langle |E_{1,2}|^2\rangle \to |E_{1,2}|^2$, and similarly for all other averaged terms. Under this condition, Eqs.~(\ref{Eq:RCircLLE}-\ref{Eq:FPCircLLE}) all reduce to the same generalized form
\begin{equation}\label{Eq:GenLLE}
    \frac{\partial E_{1,2}}{\partial t} = E_{\mathrm{in}} - \left[1 + \ii \left(-\theta + A|E_{1,2}|^2 + B|E_{2,1}|^2\right) \right]E_{1,2},
\end{equation}
\noindent where, for Eq.~\eqref{Eq:FPCircLLE}, we have recast $(3A, 3B) \rightarrow (A, B)$.

While not discussed here, we note that other coupling mechanisms are also possible. These include purely linear coupling -- for example, in inter-resonator configurations~\cite{ghosh_controlled_2024,ghosh_phase_2024,pal_linear_2024}, or in systems with significant backscattering~\cite{skryabin_hierarchy_2020}. More complex scenarios involve both linear and nonlinear couplings, such as coupled resonators supporting multiple field components~\cite{ghosh_four-field_2023}, and, additionally, one may consider coupling more than two LLEs, for instance by combining Eqs.~(\ref{Eq:RCircLLE},\ref{Eq:RCountLLE}) into a unified system~\cite{hill_multi-stage_2023-1}.
\subsection{\label{sec:SSBIntro}An Introduction to Spontaneous Symmetry Breaking}
\begin{figure*}
    \centering
    \includegraphics[width=0.775\textwidth]{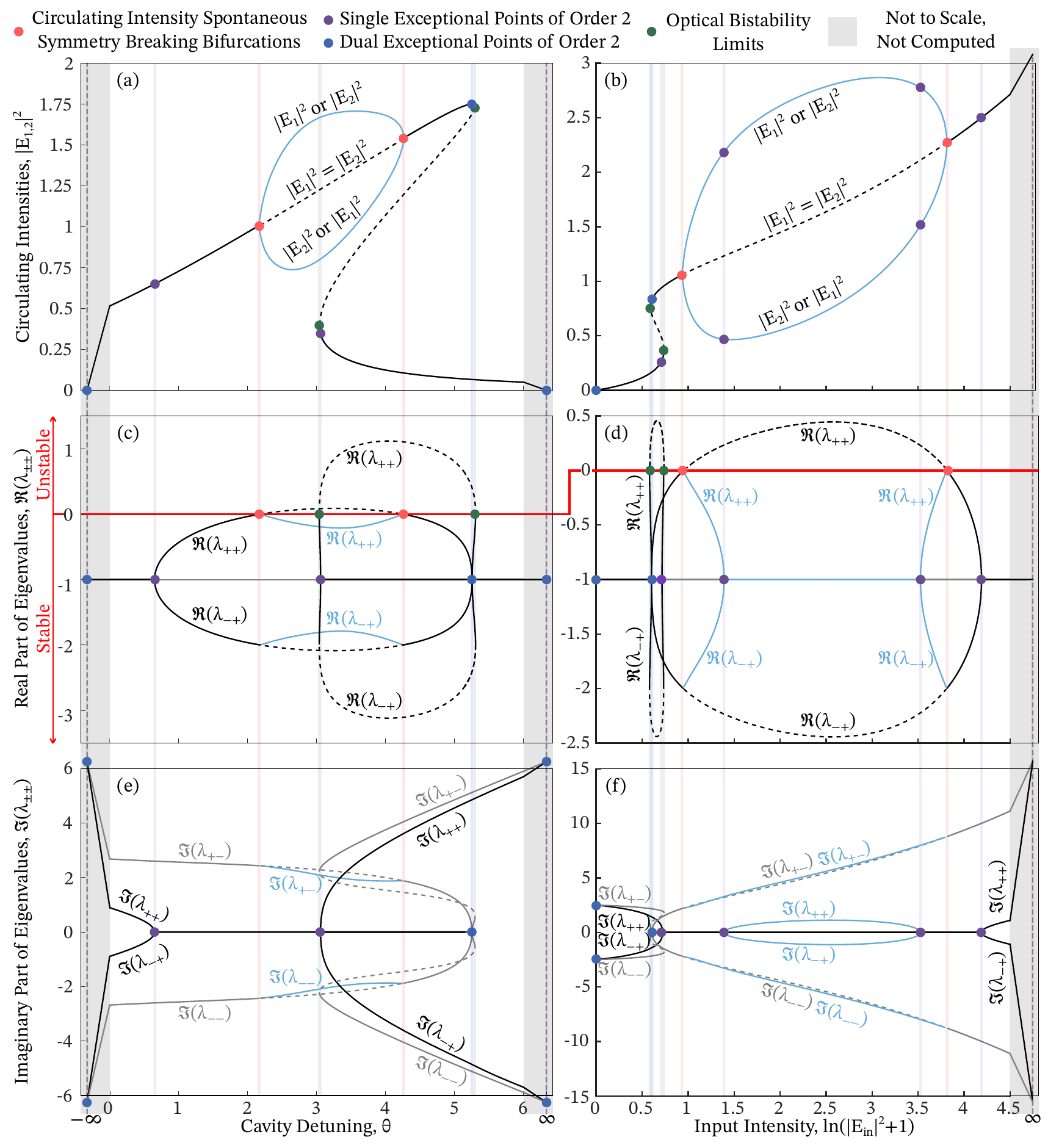}
    \caption{Solutions to Eq.~\eqref{Eq:GenLLE} are shown in panels (a) and (b) as functions of cavity detuning $\theta$ and input intensity $|E_{\mathrm{in}}|^2$, respectively. Panels (a,c,e) correspond to an input intensity of $|E_{\mathrm{in}}|^2 = 1.75$, while (b,d,f) use a detuning of $\theta = 2.5$. The solutions from (a) and (b) are inserted into Eq.~\eqref{Eq:EVs} to compute the eigenvalues $\lambda_{\pm_1\pm_2}$. The real and imaginary parts of these eigenvalues are plotted in panels (c,d) and (e,f), respectively.
    Red dots mark SSB bifurcations, purple dots denote single EP$2$s, blue dots indicate dual EP$2$s, and green dots show the limits of optical bistability. The stability of the solutions in (a) and (b) can be inferred from the real parts of the corresponding eigenvalues in (c) and (d): when $\Re(\lambda) > 0$ for any $\lambda \equiv \lambda_{\pm_1\pm_2}$, the associated solution is unstable, indicated by dashed lines across the figure.
    Eigenvalue branches that deviate from $\Re(\lambda) = -1$ in (c,d) and from $\Im(\lambda) = 0$ in (e,f) are labeled. All lines associated with symmetry-broken solutions are shown in blue. Symmetric solutions are colored black for $\Re(\lambda_{\pm+})$ and gray for~$\Re(\lambda_{\pm-})$.}
    \label{fig:EigenPlot}
\end{figure*}
In general, SSB occurs at a point in parameter space where two previously equal properties of a system -- i.e., a symmetric state -- suddenly become unequal following a small change in system parameters, as illustrated in Fig.~\ref{fig:EP_Introduction}(a,c). SSB is a central concept in many areas of physics; for example, it underpins the Higgs mechanism in particle physics~\cite{bernstein_spontaneous_1974} and describes phase transitions such as superconductivity and magnetism in condensed matter systems~\cite{baskaran_gauge_1988}. Here, of course, we focus on the SSB of light field components within Kerr resonators.

To investigate the connection between SSB and EPs, we focus on the stationary states of Eq.~\eqref{Eq:GenLLE}, defined by $\partial_t E_{1,2} = 0$. These states satisfy
\begin{equation}\label{Eq:HSS}
   E_{\mathrm{in}}=\left[1 + \ii \left(-\theta + A|E_{1,2}|^2 + B|E_{2,1}|^2\right) \right]E_{1,2}.
\end{equation}
\noindent Multiplying both sides of Eq.~\eqref{Eq:HSS} by their respective complex conjugates yields the intensity relation
\begin{equation}\label{Eq:IntHSS}
   |E_{\mathrm{in}}|^2 = |E_{1,2}|^2 \big[1 + \left(-\theta + A|E_{1,2}|^2 + B|E_{2,1}|^2\right)^2 \big].
\end{equation}
By eliminating $|E_{2,1}|^2$ from Eqs.~\eqref{Eq:HSS} and \eqref{Eq:IntHSS}, one obtains a relation depending only on $|E_{1,2}|^2$, $\theta$, and $|E_{\mathrm{in}}|^2$. Solving this relation and varying either of the two control parameters -- $\theta$ (detuning) or $|E_{\mathrm{in}}|^2$ (input intensity) -- produces parameter scans, as exemplified in Fig.~\ref{fig:EigenPlot}(a,b).

These solutions can be categorized into two types: \emph{symmetric}, where $|E_1|^2 = |E_2|^2$ (shown as black lines), and \emph{symmetry-broken}, where $|E_1|^2 \neq |E_2|^2$ (blue lines). The points in parameter space at which the symmetric solution becomes unstable and a pair of symmetry-broken solutions emerge are known as SSB bifurcations.

Various properties of Eq.~\eqref{Eq:IntHSS} -- including the parameter limits for observing SSB, thresholds for optical bistability, and system responses to asymmetric input conditions (e.g., unequal detunings or input intensities between the two fields) -- have been extensively studied \cite{kaplan_enhancement_1981,geddes_polarisation_1994,woodley_universal_2018,hill_symmetry_2024}. Importantly, it is known that the solutions lying between the optical bistability thresholds and the SSB bifurcation points are \emph{unstable}, a fact that will be central to our later analysis.
\subsection{\label{sec:EPIntro}An Introduction to Exceptional Points}
Exceptional points are ubiquitous in non-Hermitian systems~\cite{kato_perturbation_1995,berry_physics_2004,heiss_physics_2012}, where they underpin a wide range of phenomena, including unidirectional invisibility~\cite{lin_unidirectional_2011}, negative refraction~\cite{fleury_negative_2014}, light stopping~\cite{goldzak_light_2018}, and, more broadly, the tailoring of optical response functions~\cite{miri_exceptional_2019}. An EP of order $2$ (EP$2$) is a point in parameter space at which two eigenvalues become degenerate, and the two corresponding eigenvectors also coalesce. This constitutes a genuinely non-Hermitian degeneracy.

To illustrate this, consider the coupled-mode equation~\cite{miri_exceptional_2019} $\partial_\xi\, (a_1 \, a_2)^T = - \mathrm{i} H (a_1 \, a_2)^T$, where $\xi$ is an evolution parameter (e.g., time or propagation distance), and the Hamiltonian
\begin{equation}
    \label{eq:H}
    H = \begin{pmatrix}
        \omega + \ii \gamma & \kappa \\
        \kappa & \omega - \ii \gamma
    \end{pmatrix}, 
\end{equation}
with real parameters: $\omega$ is the resonance frequency; $\gamma$ represents gain in mode $a_1$ and loss in mode $a_2$; and $\kappa$ is the coupling strength between the two modes.

Similar Hamiltonians arise in a wide range of physical systems, such as coupled optical cavities~\cite{haus_coupled-mode_1991} and waveguides~\cite{yariv_coupled-mode_1973}, Bragg gratings~\cite{gordon_pmd_2000}, nonlinear crystals~\cite{kogelnik_coupled_1969}, optomechanical resonators~\cite{aspelmeyer_cavity_2014}, and cavity quantum electrodynamics (QED) systems~\cite{fox_quantum_2006}; we will also see that this Hamiltonian is highly relevant for our problem. See Refs.~\onlinecite{miri_exceptional_2019,el-ganainy_non-hermitian_2018,ozdemir_paritytime_2019} for broader context on non-Hermitian photonics.

Assuming harmonic solutions of the form $(a_1 \, a_2)^T = (\psi_1 \, \psi_2)^T e^{-\mathrm{i} \varepsilon \xi}$ leads to the eigenvalue problem \mbox{$H \psi_\pm = \varepsilon_\pm \psi_\pm$}. The eigenvalues and corresponding (non-normalized) eigenvectors are, respectively,
\begin{subequations}
\begin{align} \label{eq:epsilonpm} \varepsilon_\pm &= \omega \pm \sqrt{\kappa^2 - \gamma^2}, \\ \label{eq:psipm} \psi_\pm &= \begin{pmatrix} \ii \gamma \pm \sqrt{\kappa^2 - \gamma^2} \\ \kappa \end{pmatrix}. \end{align}
\end{subequations}

The eigenvalues are degenerate ($\varepsilon_+ = \varepsilon_-$) when the square root vanishes, i.e., at $\kappa^2 = \gamma^2$. At the same points in parameter space, the eigenvectors coalesce ($\psi_+ = \psi_-$), which is the defining property of an EP$2$. For the special case \mbox{$\gamma = 0 = \kappa$}, the Hamiltonian is Hermitian and does not exhibit an EP$2$.

Closely related to EPs are non-Hermitian symmetries and their breaking~\cite{bernard_classification_2002-1,kawabata_symmetry_2019,delplace_symmetry-protected_2021-2,sayyad_realizing_2022,budich_symmetry-protected_2019,yoshida_symmetry-protected_2019,okugawa_topological_2019}. One of the most studied is parity-time ($\mathcal{PT}$) symmetry~\cite{bender_real_1998,bender_making_2007}: The $\mathcal{PT}$ operator acts on a state $\psi$ as $\mathcal{PT} \psi = \mathcal{A} \psi^*$, where $\mathcal{A}$ is unitary and satisfies $\mathcal{A}\mathcal{A}^* = \mathcal{I}$, where $\mathcal{I}$ is the identity. A matrix (or operator) $M$ is said to be $\mathcal{PT}$-symmetric if it satisfies $M = \mathcal{A} M^* \mathcal{A}^{-1}$. This symmetry imposes the constraint $\{\varepsilon\} = \{\varepsilon^*\}$ on the spectrum, meaning that eigenvalues are either entirely real or appear in complex-conjugate pairs~\cite{bernard_classification_2002-1}.

The Hamiltonian in Eq.~\eqref{eq:H} is $\mathcal{PT}$-symmetric with $\mathcal{A} = \sigma_x$, the Pauli-$x$ matrix. The spectral constraint is evident in Eq.~\eqref{eq:epsilonpm}: for $\kappa^2 > \gamma^2$, $\varepsilon_\pm$ are real; for $\kappa^2 < \gamma^2$, they form a complex-conjugate pair. 

Important in this context is also $\mathcal{PT}$ symmetry breaking: A system is said to be in the $\mathcal{PT}$-unbroken phase if its eigenstates are also eigenstates of the $\mathcal{PT}$ operator, and in the spontaneously broken phase otherwise. The transition between these phases occurs at an EP.

In our specific case, one finds
\begin{align}
    \mathcal{PT}\,\psi_\pm = \sigma_x \psi_\pm^* = \begin{cases}
        \phantom{-}e^{+\ii \phi_\pm} \psi_\pm, & \kappa^2 > \gamma^2, \\ -e^{-\ii \phi_\pm} \psi_\mp, & \kappa^2 < \gamma^2,
    \end{cases}
\end{align}
where $e^{\pm \ii \phi_\pm}$ are complex phase factors. Thus, $\kappa^2 > \gamma^2$ corresponds to the unbroken phase, and $\kappa^2 < \gamma^2$ to the spontaneously broken phase, with the EP$2$ at $\kappa^2 = \gamma^2$ marking the transition.

Another symmetry of interest is chiral symmetry (CS) and its generalization, quasi-chiral symmetry (qCS). A~Hamiltonian exhibits CS if it satisfies $H=-\Gamma H^\dagger \Gamma^{-1}$, where $\Gamma$ is a unitary operator with $\Gamma^2 = \mathcal{I}$, and $\dagger$ denotes the Hermitian adjoint (the combination of transposition and complex conjugation). If a symmetry holds up to an additive identity term, we call it a quasi-symmetry~\cite{grafe_correlations_2013}, analogous to passive $\mathcal{PT}$ symmetry~\cite{guo_observation_2009}. In such a quasi-symmetric system, the EPs occur at the same points in parameter space as in the strict symmetric case.

In our case, the shifted Hamiltonian $H - \omega \mathbb{1}_2$ exhibits CS with $\Gamma = \sigma_z$, where $\mathbb{1}_n$ is the $n \times n$ identity matrix and $\sigma_z$ is the Pauli-$z$ matrix. Thus, $H$ is quasi-chirally symmetric, and the corresponding spectral constraint reads $\{\varepsilon-\omega\}=\{-(\varepsilon-\omega)^*\}$.

As with $\mathcal{PT}$ symmetry, qCS can also be broken. In our example, $\kappa^2 < \gamma^2$ corresponds to the unbroken qCS phase, and $\kappa^2 > \gamma^2$ to the broken phase, again with the EP$2$ marking the transition.
\section{\label{sec:ConnectingSSB&EPs}Connecting Spontaneous Symmetry Breaking and Jacobian Exceptional Points}
To analyze both the stability of solutions to Eq.~\eqref{Eq:GenLLE} -- which gives rise to SSB -- and the occurrence of EPs in the Jacobian, we begin by deriving the Jacobian matrix and its eigenvalues. We perturb a stationary solution as $E_{1,2} \to E_{1,2} + \epsilon_{1,2}$, where $\epsilon_{1,2}$ are infinitesimal perturbations. Substituting into Eq.~\eqref{Eq:GenLLE}, assuming $\partial_t E_{1,2} = 0$ and neglecting terms quadratic in $\epsilon_{1,2}$, yields a linearized system: $\partial_t ( \epsilon_1 \, \epsilon_2 \, \epsilon_1^* \, \epsilon_2^* )^T = \mathbf{J} \cdot ( \epsilon_1 \, \epsilon_2 \, \epsilon_1^* \, \epsilon_2^*)^T$, where $\mathbf{J}$ is the Jacobian matrix, with $\epsilon_{1,2}$ and their complex conjugates~$\epsilon_{1,2}^*$ treated as independent variables.

The Jacobian has the block structure 
\begin{equation}
    \label{eq:JBlock}
    \mathbf{J} =\begin{pmatrix}
        -\mathbb{1}_2 + \ii G & K \\
        K^* & -\mathbb{1}_2 - \ii G^*
    \end{pmatrix},
\end{equation}
where the $2 \times 2$ matrices $G$ and $K$ are defined as
\begin{subequations}
\begin{align}
    G &= -
    \begin{pmatrix}
        G_{11} & B E_1 E_2^* \\
        B E_1^* E_2 & G_{22}     
    \end{pmatrix}, \\
    K &= -\ii 
    \begin{pmatrix}
        A E_1^2 & B E_1 E_2 \\
        B E_1 E_2 & A E_2^2
    \end{pmatrix},
\end{align}
\end{subequations}
with $G_{11} = 2 A |E_1|^2 + B |E_2|^2 - \theta$ and $G_{22} = 2 A |E_2|^2 + B |E_1|^2 - \theta$. Since $G_{11}, G_{22} \in \mathbb{R}$, $G$ is Hermitian.

This structure allows us to identify $\mathbf{J}$ with the non-Hermitian Hamiltonian in Eq.~\eqref{eq:H}, through the correspondences $-\mathbb{1}_2 \leftrightarrow \omega$, $G \leftrightarrow \gamma$, and $K \leftrightarrow \kappa$. The Jacobian inherits $\mathcal{PT}$ symmetry via $\mathcal{A} = \sigma_x \otimes \mathbb{1}_2$, as well as quasi-chiral symmetry (qCS) through $\Gamma = \sigma_z \otimes \mathbb{1}_2$.

To characterize the eigenvalues of $\mathbf{J}$, we follow the formalism introduced in Ref.~\onlinecite{montag_symmetry-induced_2024}. We begin by defining a shifted Jacobian $\tilde{\mathbf{J}} = \mathbf{J} + \mathbb{1}_4$, which is traceless, and introduce two real-valued invariants: $\eta=\tr(\mathbf{\tilde{J}}^2)/4$ and $\nu = \det(\mathbf{\tilde{J}})$, where $\mathrm{tr}$ and $\det$ denote the trace and determinant, respectively. Using these, the eigenvalues of $\mathbf{J}$ are expressed as:
\begin{equation}
    \label{Eq:EVs}
    \lambda_{\pm_1\pm_2}=-1 \pm_1 \sqrt{\eta\pm_2\sqrt{\eta^2-\nu}},
\end{equation}
where both $\pm$ signs can be chosen independently. The parameters $\eta$ and $\nu$ are given explicitly by
\begin{equation}
\label{eq:etanu}
    \eta = - (\alpha_1\beta_1+\alpha_2\beta_2)/2, \quad \nu = - \alpha_1\alpha_2\left(\zeta^2 - \beta_1 \beta_2\right),
\end{equation}
in agreement with Ref.~\citenum{woodley_universal_2018}, where
\begin{subequations}
\begin{align}
    \alpha_{1,2} &= \theta - \phantom{3} A|E_{1,2}|^2-B|E_{2,1}|^2, \\
    \beta_{1,2} &= \theta -3A|E_{1,2}|^2-B|E_{2,1}|^2, \\
    \zeta^2 &= 4 B^2 |E_1|^2 |E_2|^2.
\end{align}
\end{subequations}
Equation~\eqref{Eq:EVs} is used to analyze the Jacobian spectrum and identify the occurrence of EPs and instability-induced symmetry breaking.

\begin{figure}
    \centering
    \includegraphics[width=\linewidth]{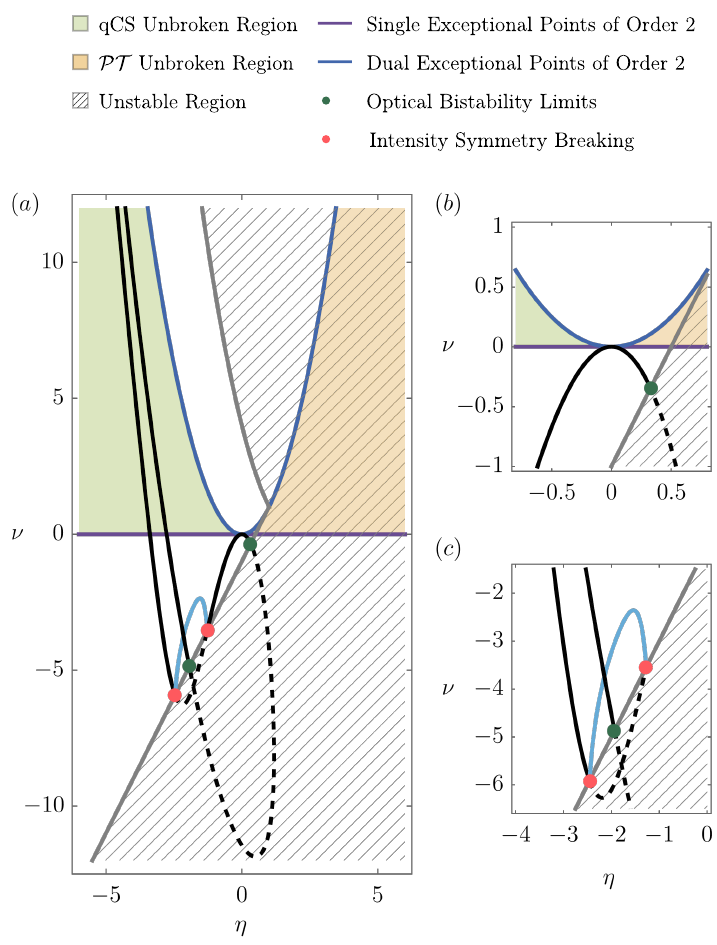}
    \caption{\textbf{Structure of the Jacobian.} (a) Full view of the Jacobian's structure in the $\eta$-$\nu$-plane, showing its EP configuration and stability regions. (b,c) Zoom-ins on the origin and the region near the SSB bifurcation, respectively. Green region: Jacobian is quasi-chiral symmetry (qCS) unbroken and $\mathcal{PT}$-broken. Orange region: $\mathcal{PT}$-unbroken and qCS-broken. White regions: both symmetries broken. The regions are separated by a curve of single EP$2$s (purple) and a curve of dual EP$2$s (dark blue). The gray line marks the stability boundary of Eq.~\eqref{Eq:GenLLE}; the gray dashed area denotes instability. Overlaid are the trajectories of the detuning scan corresponding to Fig.~\ref{fig:EigenPlot}(a,c,e). The black line represents the symmetric solution, with dashed segments indicating instability. When the trajectory enters the unstable region, the system either exhibits optical bistability (dark green dots) or undergoes SSB (red dots). The asymmetric solution is shown as the light blue curve. As $\theta \to \pm \infty$, the Jacobian asymptotically approaches the dual EP$2$ curve from within the qCS-unbroken region (green).}
    \label{fig:eta-nu}
\end{figure}

\subsection{Determining Jacobian Exceptional Points}
\label{sec:ep}
Degenerate eigenvalues of $\mathbf{J}$ arise when either the inner or outer square root in Eq.~\eqref{Eq:EVs} vanishes. As detailed in the Methods, analysis of the corresponding eigenvectors reveals that when the inner square root vanishes, the system hosts two simultaneous EP$2$s, referred to here as a dual EP$2$. In contrast, when only the outer square root vanishes, the system exhibits a single EP$2$. This leads to the following:
\begin{align}
    \label{eq:ep}
    \begin{cases}
        \text{dual EP}2, & \text{if } \nu = \eta^2, \\
        \text{single EP}2, & \text{if } \nu = 0 \text{ and } \eta \neq 0.
    \end{cases}
\end{align}
As discussed in Sec.~\ref{sec:EPIntro}, these EPs mark transitions between symmetry phases. Figure~\ref{fig:eta-nu} shows the locations of single and dual EP$2$s in the $\eta$-$\nu$-plane (purple and dark blue curves, respectively). These curves partition the plane into distinct symmetry regimes: green corresponds to the qCS-unbroken and $\mathcal{PT}$-broken phase, orange to the $\mathcal{PT}$-unbroken and qCS-broken phase, and white to both symmetries broken.
\subsection{Spontaneous Symmetry Breaking Bifurcations for Circulating Intensities}
\label{sec:ssb}
SSB of the circulating intensities arises from instabilities in the solutions to Eq.~\eqref{Eq:GenLLE}, specifically when the real part of any Jacobian eigenvalue $\lambda_{\pm_1 \pm_2}$ becomes positive. As the eigenvalue $\lambda_{++}$ always possesses the largest real part, it suffices to consider the condition $\Re(\lambda_{++}) \geq 0$. This stability criterion also predicts the onset of related phenomena such as optical bistability, as illustrated in Figs.~\ref{fig:EigenPlot} and~\ref{fig:eta-nu}. Solving this inequality yields the following condition:
\begin{equation}
    \label{eq:ssb}
    \eta \geq \begin{cases}
        2-\sqrt{\nu},  & \nu \geq 1,  \\
        (1+\nu)/2, & \nu \leq 1,
    \end{cases}
\end{equation}
with equality marking the boundary of stability. This boundary is shown in Fig.~\ref{fig:eta-nu} as a gray curve, with the dashed region indicating instability.
\subsection{SSB and Jacobian EPs are Different, But \ldots}
\label{sec:ssbnotep}
The algebraic conditions for Jacobian EPs (Eq.~\eqref{eq:ep}) and intensity SSB (Eq.~\eqref{eq:ssb}) are distinct. As a consequence, these phenomena generally occur at different locations in parameter space. However, a closer inspection of the eigenvalue trajectories in Fig.~\ref{fig:EigenPlot}(c,d) reveals a deeper underlying connection between them.

Both the detuning and intensity scans begin and end in a regime where $\Re(\lambda_{\pm_1\pm_2}) = -1$, corresponding to the qCS-unbroken phase of the Jacobian. Therefore, for SSB to occur, the system must pass through an EP and undergo qCS breaking.

This observation is generic for Eq.~\eqref{Eq:GenLLE}. In the limits $\theta \to \pm \infty$ or $|E_{\text{in}}|^2 \to 0,\infty$, the symmetric solution describes the system for $B < 3A$~\cite{hill_effects_2020} (see Methods for $B > 3A$). In this case, the relation between input and circulating intensities is
\begin{align}
    \label{eq:symmetric}
    I = P \left[ 1+ \left(\theta -(A+B) P \right)^2\right],
\end{align}
where $I = |E_{\text{in}}|^2$ and $P = |E_1|^2 = |E_2|^2$.

Solving Eq.~\eqref{eq:symmetric} for $\theta$ gives
\begin{equation}
    \label{eq:thetapm}
    \theta_\pm(P,I) = (A+B) P \pm \sqrt{I/P-1},
\end{equation}
from which limits can be analyzed as $P \to 0^+$ or $P \to \infty$.

For detuning scans, taking $P \to 0^+$ and inserting into Eq.~\eqref{eq:etanu} yields $\lim_{P \to 0^+} (\eta, \nu) = (-\infty, \infty)$, placing the system in the second quadrant of the $\eta$-$\nu$-plane. We define $\nu = \eta^2 - d$ with
\begin{equation}
    \label{eq:d}
    d(\theta,P) = 4 B^2 P^2 \left[ \theta - (A+B) P\right]^2.
\end{equation}
Eliminating $\theta$ using Eq.~\eqref{eq:thetapm} gives $d = 4 B^2 (I - P)P$. As $P \to 0^+$, $d \to 0$, so the system asymptotically approaches the dual EP$2$ line from below, i.e., from within the qCS-unbroken region. A similar argument holds for intensity scans: the limits $I \to 0$ and $I \to \infty$ correspond to $P \to 0$ and $P \to \infty$, respectively. In both cases, the system lies within the qCS-unbroken regime.

Taken together, these results show that crossing a Jacobian EP is a necessary condition for the onset of instability. Jacobian EPs therefore act as structural precursors to spontaneous symmetry breaking.
\section{\label{sec:Conclusion}Conclusion}
\vspace{-0.2cm}
\noindent In this paper, we have examined a nuanced relationship between SSB and EPs in nonlinear optical systems, with a particular focus on Kerr resonators.

Through a detailed theoretical analysis of three experimentally relevant nonlinear systems, we have shown that SSB and Jacobian EPs generally occur at distinct locations in parameter space. However, the emergence of Jacobian EPs is a necessary precursor to the onset of SSB. This finding challenges the often-assumed equivalence between EPs and symmetry-breaking phenomena and highlights the importance of distinguishing between different classes of EPs in both theory and application.

Looking forward, this framework can be extended to other classes of nonlinear systems. Beyond that, we are currently compiling a complementary study that systematically identifies which classes of EPs accurately predict and explain the onset of SSB~\cite{bai_nonlinear_2023,bai_observation_2024,yoshida_exceptional_2025,fang_exceptional_2024}.

These insights have significant implications for the design and control of photonic devices. A clearer understanding of the interplay between EPs and SSB enables more precise tuning of nonlinear optical behavior and opens new avenues for the development of symmetry-driven functionalities in resonant photonic platforms.
\vspace{-0.2cm}
\section*{Acknowledgments}
\vspace{-0.2cm}
We thank Alexander Felski, George Ghalanos, Kyle Kawagoe, Jona Kayser, and Anton Montag for insightful discussions. We also acknowledge early, separate, and unpublished work by Michael T.M. Woodley exploring the connection between Jacobian EPs and SSB in counter-propagating light, which came to our attention late in the development of this project. L.H. acknowledges funding from the Max Planck Society (MPG) and the Centre national de la recherche scientifique (CNRS). J.T.G. and F.K.K. acknowledge support from the MPG Lise Meitner Excellence Program 2.0. L.H., J.T.G., and F.K.K. also acknowledge funding from the European Union's ERC Starting Grant “NTopQuant” (101116680). L.H. and P.D'H. acknowledge support through the ERC Starting Grant “CounterLight” (756966). The views expressed are those of the authors and do not necessarily reflect those of the EU or the ERC. Neither the EU nor the granting authority can be held responsible for them.
\vspace{-0.2cm}
\section*{Author Contributions}
\vspace{-0.2cm}
L.H. and J.T.G. carried out the majority of the calculations and theoretical work and should be considered co-first authors. They were supported by A.G. and J.F., who contributed to discussions and the analysis of results. F.K.K. and P.D'H. initiated the project and provided ongoing guidance, engaging in discussions with all authors. L.H. and J.T.G. prepared the manuscript with input from all co-authors.
\section*{Methods}
\subsection{Denormalizing the LLE}
For both Kerr ring and Fabry-Pérot resonators, the normalized Lugiato-Lefever equation (LLE) can be mapped to experimental parameters via the following transformations~\cite{leo_temporal_2010}:
\begin{equation}\label{Eq:Trans}
    \begin{aligned}
    t\rightarrow\frac{\alpha t}{\tau_\textrm{R}}, && \tau\rightarrow\tau\sqrt{\frac{2\alpha}{|\beta_2|L}}, && E_{1,2}\rightarrow E_{1,2}\sqrt{\frac{\gamma L}{\alpha}},
    \end{aligned}
\end{equation}
together with the substitutions:
\begin{equation}\label{Eq:Trans2}
    E_{\mathrm{in}}=S_{\mathrm{in}}\sqrt{\frac{\gamma L\theta}{\alpha^3}}, \qquad \theta=\frac{2m\pi-\phi_0}{\alpha},
\end{equation}
where $S_{\mathrm{in}}$ is the amplitude of the driving field incident on the input coupler. The loss coefficient is given by $\alpha = \pi / \mathcal{F}$, equal to half the fractional intracavity power loss per round trip, where $\mathcal{F}$ is the cavity finesse. The round-trip time is denoted by $\tau_\mathrm{R}$, $L$ is the physical cavity length, $\gamma$ is the Kerr nonlinear coefficient, $\beta_2$ is the second-order dispersion parameter, $m$ is the order of the nearest resonance, and $\phi_0$ is the round-trip phase shift.

The output field is given by:
\begin{equation}\label{Eq:Output}
    E_{\textrm{out}} = \sqrt{\frac{\alpha\theta}{\gamma L}}(E-\kappa E_{\mathrm{in}}),
\end{equation}
where $\kappa=\alpha/\theta$.

In some literature, cavity loss is expressed in terms of the cavity linewidth $\kappa_{\mathrm{tot}}$ (in Hz), which is related to the photon lifetime by $\tau_{\mathrm{ph}} = 1/\kappa_{\mathrm{tot}}$. The loss coefficient $\alpha$ relates to the linewidth via: $\alpha = \pi \kappa_{\text{tot}} / \text{FSR}$, where FSR is the cavity's free spectral range. The cavity detuning $\Delta \omega = \omega_l - \omega_{\mathrm{res}}$ is defined as the frequency difference between the laser pump $\omega_l$ and the nearest cavity resonance $\omega_{\mathrm{res}}$, and is related to $\theta$ by: $\theta = \Delta \omega / \text{FSR}$.

The nonlinear parameter is given by: $\gamma = \omega_0 n_2 / c A_{\text{eff}}$, where $n_2 = \varepsilon_0 c n \bar{n}_2 / 2$ is the nonlinear refractive index (with $\bar{n}_2$ in units of $\mathrm{W}^{-1}\mathrm{m}^2$), $n$ is the linear refractive index, $\varepsilon_0$ is the vacuum permittivity, $c$ is the speed of light in vacuum, and $A_{\mathrm{eff}}$ is the effective mode area. The units of $\gamma$ are $\mathrm{W}^{-1}\mathrm{m}^{-1}$.

In ring-resonator systems, $\gamma$ is sometimes defined through the four-wave mixing gain: $g_0 = c^2 \hbar \omega_0 \gamma / 2 \pi n^2 R$, where $R$ is the resonator radius and $\hbar$ is the reduced Planck constant.
\subsection{Properties of the Jacobian}
To determine both the symmetry phases of the Jacobian and its stability, we define the inner square root of Eq.~\eqref{Eq:EVs} as
\begin{equation}
    S = \sqrt{\eta^2-\nu},
\end{equation}
such that the eigenvalues of the Jacobian can be written as $\lambda_{\pm_1\pm_2}=-1 \pm_1 \sqrt{\eta \pm_2 S}$.
\subsubsection{Symmetry Phases of the Jacobian}
To identify the symmetry phases, we analyze the regions in the $\eta$-$\nu$-plane shown in Fig.~\ref{fig:eta-nu}:
\begin{enumerate}[label=\Roman*.]
    \item $\nu > 0$ and $\eta > \sqrt{\nu}$: $S$ is real and $\eta > S$, so all $\lambda_{\pm_1\pm_2}$ are real. The Jacobian is in the $\mathcal{PT}$-unbroken phase.
    \item $\nu > 0$ and $-\sqrt{\nu} < \eta < \sqrt{\nu}$: $S$ is purely imaginary, making $\eta \pm S$ complex. Hence, all eigenvalues are complex and both $\mathcal{PT}$ and qCS symmetries are broken.
    \item $\nu > 0$ and $\eta < -\sqrt{\nu}$: $S$ is real and $\eta < S$, which implies $\lambda_{\pm_1\pm_2} \in \{-1 + \ii \mathbb{R}\}$. The Jacobian lies in the qCS-unbroken phase.
    \item $\nu < 0$: $S$ is real and $\eta < S$, yielding $\lambda_{\pm+} \in \mathbb{R}$ and $\lambda_{\pm-} \in \{-1 + \ii \mathbb{R}\}$. Thus, both symmetries are broken.
\end{enumerate}
\subsubsection{Stability from the Jacobian}

We now analyze stability using the same regions in the previous section and Fig.~\ref{fig:eta-nu} to derive Eq.~\eqref{eq:ssb}. As $\lambda_{++}$ always has the largest real part, we consider the condition $\Re(\lambda_{++}) \geq 0$, which is equivalent to
\begin{equation}
    \label{eq:ineq}
    \Re\left( \sqrt{\eta+S} \right) \geq 1.
\end{equation}
In Region III (qCS-unbroken), all eigenvalues satisfy $\Re(\lambda) = -1$, so this region is entirely stable.

In Regions I and IV, where $S$ is real and $\eta + S \geq 0$, Eq.~\eqref{eq:ineq} becomes: $\sqrt{\eta+S}\geq 1$. This can be rearranged into: $\sqrt{\eta^2-\nu}\geq 1 - \eta$. If $\eta \leq 1$, this inequality can be solved for $\nu$. If $\eta > 1$, the inequality is trivially satisfied, since the left-hand side is always real and non-negative. This yields:
\begin{equation}
    \label{eq:SrealIneq}
    \nu \leq \begin{cases}
    2 \eta - 1,    & \text{if }\eta < 1, \\
    \eta^2,    & \text{if } \eta > 1.
    \end{cases}
\end{equation}
In Region II, where $S$ is imaginary, we use the identity~\cite{abramowitz_handbook_1964}:
\begin{equation*}
    \Re\left({\sqrt{a + \ii b}}\right)=\sqrt{\frac{\sqrt{a^2+b^2}+a}{2}},
\end{equation*}
for real $a, b$. Setting $a = \eta$ and $b = -\ii \sqrt{\eta^2 - \nu}$, so $b^2 = -(\eta^2 - \nu)$, and applying to Eq.~\eqref{eq:ineq}, we find
\begin{align}
    \label{eq:ScomplexIneq}
    \nu \geq (2 - \eta)^2, \qquad \text{if $\nu \geq \eta^2$}.
\end{align}
Combining Eq.~\eqref{eq:SrealIneq} and Eq.~\eqref{eq:ScomplexIneq} recovers the instability boundary defined in Eq.~\eqref{eq:ssb}.
\subsubsection{Limits of the parameter scans}
\vspace{-0.5cm}
In the main text, we focus on the case where the symmetry-broken region is bounded, i.e., when \mbox{$B < 3A$}.
For intensity scans with $B > 3A$, however, symmetry is not restored at high input power, and the symmetric solution (Eq.~\eqref{eq:symmetric}) no longer describes the limit $I \to \infty$~\cite{hill_effects_2020}. In this regime, the system remains in a symmetry-broken state. To analyze the asymptotic behavior, we consider the two symmetry-broken solutions $P_1(\theta, I)$ and $P_2(\theta, I)$, with $P_{1,2} = |E_{1,2}|^2$, and extend the approach of the main text by evaluating $d \equiv d(\theta, P_1, P_2)$.

Assuming $P_1$ and $P_2$ grow with the same asymptotic scaling, the leading-order contribution to $d$ as \mbox{$P_1, P_2 \to \infty$} is

\begin{widetext}
    \vspace{-0.3cm}
    \begin{equation}
    d(\theta,P_1,P_2)
    = \frac{1}{4} \left(B^2-3 A^2\right)^2
   \left(P_1^4+P_2^4\right) + \frac{P_1^2
   P_2^2}{2} \left[13 A^2 B^2+A^2 \left(B^2-9 A^2\right)+7 B^4\right] + 4 A B^3 P_1 P_2
   \left(P_1^2+P_2^2\right)
\end{equation}
    \vspace{-0.5cm}
\end{widetext}

As in the symmetric case, all terms are positive for \mbox{$B > 3A$}. Therefore, intensity scans in this regime remain in the qCS-unbroken phase at large $P_1$ and $P_2$. \newpage

We conclude that, for $B > 3A$, the dual EP$2$ line is not approached in the large-intensity limit of the symmetry-broken branch.
\vspace{-0.7cm}

\bibliography{references}
\end{document}